# High-Pressure Monoclinic-Monoclinic Transition in Ferguson ite-type HoNbO$_4$

A. B. Garg[1,2], D. Errandonea[3], P. Rodríguez-Hernández[4], and A. Muñoz[4]

1. High Pressure and Synchrotron Radiation Physics Division, Bhabha Atomic Research Centre, Mumbai 400085, India

2. Homi Bhabha National Institute, Anushaktinagar, Mumbai 400094, India

3. Departamento de Física Aplicada-ICMUV, Universidad de Valencia, Dr. Moliner 50, Burjassot, 46100 Valencia, Spain

4. Departamento de Física, Instituto de Materiales y Nanotecnología, Universidad de La Laguna, La Laguna 38205, Tenerife, Spain

*e-mail: daniel.errandonea@uv.es

**Abstract**

In this paper we perform a high-pressure study of fergusonite-type HoNbO$_4$. Powder x-ray diffraction experiments and *ab initio* density-functional theory simulations provide evidence of a phase transition at 18.9(1.1) GPa from the monoclinic fergusonite-type structure (space group I2/a) to another monoclinic polymorph described by space group P2$_1$/c. The phase transition is reversible and the high-pressure structural behavior is different than the one previously observed in related niobates. The high-pressure phase remains stable up to 29 GPa. The observed transition involves a change in the Nb coordination number from 4 to 6, and it is driven by mechanical instabilities. We have determined the pressure dependence of unit-cell parameters of both phases and calculated their room-temperature equation of state. For the fergusonite-phase we have also obtained the isothermal compressibility tensor. In addition to the high-pressure studies, we report ambient-pressure Raman and infrared spectroscopy measurements. We have been able to identify all the active modes of fergusonite-type HoNbO$_4$, which have been assigned based upon density-functional theory calculations. These simulations also provide the elastic constants of the different structures and the pressure dependence of the Raman and infrared modes of the two phases of HoNbO$_4$.

**INTRODUCTION**

Lanthanide niobates with formula $RNbO_4$ (where R is a trivalent lanthanide) have recently attracted broad attention because of their promising properties as multi-functional materials [1]. Most of the studies on these compounds have been focused on their structural, vibrational, dielectric, luminescent, and magnetic properties [2]. The members of this family are known to have a monoclinic crystal structure, described by space group $I2/a$, usually named as fergusonite. The crystal structure has been accurately determined by neutron diffraction [3]. In Fig. 1, it can be seen that this structure is composed by chains of alternating $RO_8$ and $NbO_4$ polyhedra [3]. It is known that at high temperature, $RNbO_4$ compounds undergo a ferroelastic phase transition [4], having the high-temperature phase a tetragonal scheelite-type structure (space group $I4_1/a$) [5, 6]. $RNbO_4$ have also been studied under high-pressure (HP) conditions. Such studies have proved to be an extremely efficient tool to test the structural stability and determine the mechanical properties of ternary oxides [7]. However, in comparison with related oxides, like tunsgtates and vanadates [8, 9], orthoniobates have been much less studied. In addition, there are apparent contradictions in the reported results as described in detail in next paragraphs.

The first HP study on $RNbO_4$ compounds was carried out in $LaNbO_4$ more than three decades ago [10]. In this study, the maximum pressure achieved was 3.3 GPa and it was found that the monoclinic distortion of $LaNbO_4$ increases with increasing pressure. The second work on orthoniobates under compression was reported at the beginning of the 21[st] century [11]. In this work, it was stated that a metastable orthorhombic polymorph was obtained at 8 GPa and 1570 K. More recently, studies have been carried out in $EuNbO_4$ [12]. In them, subtle changes in the Raman and luminescent spectra have been identified as the onset of a phase transition at 7.3 GPa. The proposed HP phase [12] is in apparent contradiction with the results reported for $LaNbO_4$ [10, 11]. A similar phase transition to that of $EuNbO_4$ has been found for Eu-doped $GdNbO_4$ at 6.2 GPa [13]. However, both x-ray diffraction (XRD) and Raman experiments, made in un-doped $GdNbO_4$, showed that there is no phase transition in this niobate up to 25 GPa [14]. To add more complexity to the puzzle, in wolframite-type $InNbO_4$, an isostructural phase transition was discovered at 10.3 GPa [15, 16]. All the facts described above indicate the need of performing additional HP studies on $RNbO_4$ compounds. In particular, for late lanthanides compounds, with a rare earth having a nearly filled 4*f* electron shell (e.g. Ho, Er, and Tm); which have not been studied yet. This is important, because the *f* electron occupancy plays a fundamental role in the HP behavior of lanthanide compounds [9, 17].

Here we investigate the behavior of HoNbO$_4$ under HP conditions (Ho has a $4f^{11}6s^2$ electronic configuration) by means of *in-situ* XRD and density-functional theory (DFT) simulations up to 30 GPa. We will present evidence of a first-order reversible phase transition at 18.9(1.1) GPa. The crystal structure of the HP phase has been assigned to a monoclinic structure described by space group P2$_1$/c. According to calculations, the transition is driven by mechanical instabilities induced by pressure in the fergusonite structure. We will also report the room-temperature equations of state (EOS) and compressibility tensor of HoNbO$_4$. *Ab initio* calculations also provide information on the pressure dependence and assignment of the Raman and infrared (IR) modes. The good agreement with Raman and IR measurements at ambient conditions support the phonon calculations at high pressure.

**EXPERIMENTAL DETAILS**

HoNbO$_4$ was synthesized by solid-state reaction from pre-dried Gd$_2$O$_3$ and Nb$_2$O$_5$ (purity > 99.9%). Powder x-ray diffraction collected using a rotating-anode generator ($\lambda$ = 0.7107 Å) and a MAR345 area detector confirmed the single phase formation of the compound with a fergusonite-type structure. The unit-cell parameters obtained at ambient conditions are *a* = 5.082(5) Å, *b* = 10.954(9) Å, *c* = 5.310(5) Å, and *b* = 94.76(9)°, which agree well with the literature [3, 18, 19]. Notice that in the literature a setting in which a > c is used. However, this setting (with a > c) differs from the standard setting recommended by the International Union of Crystallography (with a < c). So we have used the standard setting to describe the crystal structure of fergusonite-type HoNbO$_4$.

High-pressure XRD experiments have been carried out at room-temperature up to 29.2 GPa using 4:1 methanol–ethanol as pressure-transmitting medium [20, 21]. Angle-dispersive XRD measurements have been performed employing a diamond-anvil cell. The size of the diamond culets was 400 μm. The sample was loaded in a 150 μm hole drilled in a stainless-steel gasket, pre-indented to a thickness of 50 μm. Sample loading was done carefully avoiding sample bridging between diamonds [22]. Pressure has been determined, with an error smaller than 0.1 GPa, using the EOS of silver as pressure scale [23]. HP-XRD measurements have been done at the EXPRESS beamline of Elettra synchrotron using a monochromatic wavelength of 0.5997 Å. The XRD patterns have been collected using a MAR345 image-plate detector. Structural analysis has been performed using Fullprof [24] and PowderCell [25].

The Raman spectrum was measured in backscattering configuration using a TRH 1000 Jobin-Yvon spectrometer in combination with a thermoelectric-cooled CCD detector and a 632.8 nm laser with a 10 mW incident power on the sample. The spectral resolution was better than 2 cm$^{-1}$. An infrared absorption spectrum of the sample was recorded on a JASCO Fourier-transform IR spectrometer in transmission mode.

**SIMULATIONS DETAILS**

The influence of pressure on the crystal structure and on the mechanical and vibrational properties of $HoNbO_4$ has been also analyzed performing *ab initio* simulations. The study was carried out on the framework of the density functional theory (DFT) [26]. The Vienna *Ab initio* Simulation Package (VASP) [27] was employed and the pseudopotentials were generated with the projector-augmented wave scheme (PAW) [28] including the *f* electrons into the core. This has proven to give good results in the study of $RMO_4$ compounds [29]. Due to presence of oxygen in the studied compound, to ensure accurate results, the set of plane waves was extended up to a 530 eV cutoff energy. The exchange-correlation energy was expressed by means of the Generalized-Gradient Approximation (GGA) with the AM05 [30] prescription. Dense grids of Monkhorst-Pack [31] k-special points, (4x4x6) for the low-pressure (LP) phase and (6x2x6) for the HP phase, were employed to perform the integrations on the Brillouin zone (BZ). This procedure achieves very high convergences of 1 meV per formula unit in the total energy. By minimizing the forces on the atoms (lower than 2meV/Å per atom) and the deviation of the stress tensor from a diagonal hydrostatic form (less than 0.1 GPa), all the structural parameters of both phases were optimized at selected volumes. The resulting sets of energies and volumes (*E*, *V*) were fitted with an equation of state to evaluate the equilibrium volume ($V_0$), the bulk modulus ($B_0$), and its pressure derivative ($B_0'$). The simulations were performed at zero temperature, and the stable structures and the transition pressures were obtained by analyzing the enthalpy-pressure curves. This method has been successfully applied to the study of phase stability, and to accurately describe the structural properties of semiconductors under high pressure [32].

In order to carry out the study of the mechanical properties of HoNbO$_4$ at the low- and high-pressure, the elastic constants were evaluated employing the method implemented in the VASP code: the ground state and fully optimized structures were strained in different directions taking into account their symmetry [33]. To study the Raman and infrared phonons, lattice-dynamics calculations were performed at the zone center ($\Gamma$ point) of the BZ using the direct method [34]. The diagonalization of the dynamical matrix provides the frequencies of the normal modes. Furthermore, these calculations allow to identify the symmetry and eigenvectors of the phonon modes at the $\Gamma$ point.

**RESULTS AND DISCUSSION**

In Fig. 2 we show a selection of powder XRD measured at different pressures. In addition to the peaks from the sample we detected the peaks from Ag, which was used as pressure marker. The (111) peak of Ag, which was used for pressure determination, is denoted with an asterisk in the figure. We found that up to 17.8 GPa all diffraction peaks of the sample can be assigned to the fergusonite-type structure. The results of the Rietveld refinement at 0.2 GPa are shown in the figure. The obtained unit-cell parameters are: $a$ = 5.082(5) Å, $b$ = 10.954(9) Å, $c$ = 5.310(5) Å, and $\beta$ = 94.76(9)º. The R-values of the refinement are $R_p$ = 3.09% and $R_{wp}$ = 5.54%. Similar quality of refinements has been obtained up to 17.8 GPa. For the refinements we used the atomic positions reported by Tsunekawa et *al*. [3] at ambient pressure because we found that in case of refining these positions the changes in the coordinates induced by pressure were smaller than their uncertainties.

In the first compression step beyond 17.8 GPa, at 19.9 GPa, we observed subtle changes in the XRD pattern. As a consequence, the patterns cannot be explained by the fergusonite structure. The most noticeable modifications are the gradual merging of the two most intense peaks around 9-10º and several changes observed at low angles. To facilitate the identification of these changes we show in Fig. 3, two zooms of different low-angle regions. In the 5-7º region, there is an extra peak denoted by an asterisk which appears at 19.9 GPa. In the 8.8-10.4º region, it can be seen the clear splitting of one peak. We consider these facts as the evidence of the onset of a phase transition. We estimate the transition pressure, 18.9(1.1) GPa, as the average pressure corresponding to the last XRD pattern of the low-pressure phase and the first pattern of the HP phase. We assume as the error for the transition pressure the standard deviation. The HP phase is stable up to 29.0 GPa and upon decompression the transition is reversible as shown in Fig. 2 by the pattern measured at 0.2 GPa after pressure release, which can undoubtedly be identified to the fergusonite structure.

With the help of DFT calculations we have found a different monoclinic structure that can explain the XRD measurements from 19.9 to 29.0 GPa. The HP structure belongs to space group $P2_1/c$. In Fig. 2 the results of the Rietveld refinements at 24.9 GPa show that the proposed crystal structure can satisfactory explain the results obtained for the HP phase. The refined unit-cell parameters at 24.9 GPa are $a = 4.96(2)$ Å, $b = 10.40(4)$ Å, $c = 5.15(2)$ Å, and $\beta = 96.3(3)°$. The R-values of the refinement are $R_p = 5.49\%$ and $R_{wp} = 8.05\%$. In this refinement we have used the DFT calculated atomic positions, which were not refined, and are given in Table 1. In Fig. 1 we compare the proposed HP structure with the LP fergusonite structure. The most noticeable change after the transition is the increase of the coordination number of Nb, which is four in fergusonite, but six in the HP phase, resembling the Nb coordination in diniobium hexa-oxides [35]. The transition also implies a change in the way that polyhedra are interconnected. Interestingly, a similar structure has been found as post-fergsuonite phase under HP in $Eu_{0.1}Bi_{0.9}VO_4$ [36], $LiYF_4$ [37], and $NdTaO_4$ [38], which supports the proposed crystal structure of HP phase as a likely structure.

In order to support the conclusions obtained from XRD experiments we have performed total-energy DFT calculations. These *ab initio* simulations support fergusonite structure as the stable LP phase of $HoNbO_4$. The calculated crystal lattice parameters and atomic positions at ambient conditions are given in Table 1. The agreement with the known crystal structure [3] is very good. As pressure increases we found that the fergusonite structure and the proposed HP phase have the same enthalpy, taking into account the precision of the calculations. However, after performing elastic constant calculations we found that the generalized Born stability criteria [39, 40] are violated by the fergusonite structure at 19.5 GPa. In contrast these stability criteria are satisfied by the HP phase. As consequence, according to calculations, the transition from fergusonite to the HP monoclinic structure takes place at 19.5 GPa in $HoNbO_4$, in good agreement with the experimental finding. The transition occurs because the fergusonite structure becomes mechanically unstable. The calculated crystal lattice parameters and atomic positions for the monoclinic HP phase of $HoNbO_4$ are reported in Table 1. This structure is very similar to the one obtained from XRD experiments and the theoretical transition pressure agrees with the experimental one. Thus, we can conclude that rare-earth niobates with a late lanthanide (like $HoNbO_4$) follow a different structural sequence than other niobates of the same family [14], in analogy to the behavior of $RVO_4$ vanadates [9].

In addition to analyzing the structural stability, elastic constant calculations also give information on the elastic properties of the two phases of $HoNbO_4$. Moreover, the elastic constants allow to determine interesting elastic moduli for practical applications of this material. The calculated values of the thirteen independent elastic constants for each phase are summarized in Table 2. For comparison, the table also includes previous results for fergusonite $LaNbO_4$ [41] and $YTaO_4$ [42]. The three compounds with the fergusonite structure have qualitative similitudes on the values of the elastic constants. In particular, our values for fergusonite $HoNbO_4$ compare better with $LaNbO_4$ than with $YTaO_4$. On the other hand, the comparison between the three compounds suggest that $C_{35}$ has been previously reported with the wrong sign for $LaNbO_4$ and that $C_{35}$ has been underestimated (see Table 2). Further examination of the elastic constants indicates that in $HoNbO_4$ the bonding in the [001] direction is stiffer than that in [010] and [100] directions for the fergusonite phase ($C_{33}$> $C_{11}$> $C_{22}$) indicating that the fergusonite phase is weakly anisotropic, being the *b*-axis ([010] direction) the most compressible one. A similar anisotropy is found in the HP phase where also the *b*-axis is the most compressible axis. Regarding, the HP phase, it can be seen that the values of $C_{11}$, $C_{22}$, $C_{33}$, $C_{12}$, $C_{13}$, $C_{23}$ are considerably larger than in fergusonite phase, being the differences for the other constants smaller when the two phases are compared. Therefore, transverse shear waves, corresponding to the constants $C_{44}$, $C_{55}$, and $C_{66}$, are less affected by the phase transition than longitudinal sound waves, corresponding to the constants $C_{11}$, $C_{22}$, and $C_{33}$.

From the calculated elastic constants, we have determined the Poisson's ratio (ν) and the bulk ($B_0$), shear (G), and Young (E) moduli for the two phases of $HoNbO_4$. We have computed their values using the Voigt [43], Reuss [44], and Hill [45] approximations. The obtained values are summarized in Table 3. These parameters are directly related to the response of HoNbO$_4$ to stresses, being the relevant moduli for practical applications. They are comparable to the same moduli in related ternary oxides like vanadates and phosphates [46, 47]. On the other hand, the two phases have values of the Poisson's ratio larger than 0.30. This indicates that the inter-atomic forces are predominantly central ($v > 0.25$) and that ionic bonding is predominant against covalent bonding [48]. Another relevant parameter for describing the macroscopic properties of a material is the $B_0/G$ ratio. According to Pugh [49], if this ratio is larger (smaller) than 1.75 the material behaves in a ductile (brittle) manner. Therefore, HoNbO$_4$ behaves as a ductile material as can be seen in Table 3. Macroscopic elastic anisotropy is another property of great importance in engineering. Basically, this parameter correlates to the possibility of inducing micro-cracks in the materials under stress [50]. Anisotropy can be quantified for all types of crystals with the universal elastic anisotropy index ($A_U$) defined by Ranganathan and Ostoja-Starzewski [51]. Using the equation proposed by them we have calculated $A_U$. The obtained values are given in Table 3. In the low-pressure fergusonite phase, this value is close to 0, therefore elastic anisotropy is moderate. In the HP phase, the elastic anisotropy is considerable larger (see Table 3). Hardness is also a relevant parameter which is related to the elastic and plastic properties. We have calculated the Vickers hardness ($H_V$) using the elastic moduli summarized in Table 3 (Hill approximation) and the phenomenological equation proposed by Tian *et al.* [52]. The results are also shown in Table 3. The low-pressure phase is harder than the high-pressure phase, with a hardness of approximately 7 GPa, which is equivalent to 700 Vickers hardness.

From the refinement of the XRD patterns measured at different pressures we have obtained the pressure dependence of the unit-cell parameters and volume of the two phases of HoNbO$_4$. The results are shown in Fig. 4 where they are compared with DFT calculations. For the unique *b*-axis of the structure we have plotted *b*/2 to facilitate the comparison with the other axes. Calculations qualitatively agree with the experiments, with a tendency to underestimate the unit-cell parameters and volume. For the fergusonite phase, the decrease of the volume is slightly larger in the calculations than in the experiments. On the other hand, calculations predict a more pronounced non-linear behavior for the β angle than experiments. However, both experiments and calculations give a similar behavior under compression with the crystal anisotropy decreasing with pressure. At the phase transition, both calculations and experiments do not predict any detectable volume change, however, they found a discontinuity for the β angle. This angle increases with pressure in the HP phase while it decreases with pressure in the low-pressure fergusonite phase. Apparently, there are no detectable changes in the volume compressibility at the phase transition. This and the fact that there are no volume changes (plus the reversibility of the transition) would suggest that the transition might be a second-order transition [53]. However, the changes observed in coordination polyhedra and the increase of the Nb coordination associated to the transition, are an evidence that the transition is indeed a first-order transformation [54]. Regarding the axial compressibility, we found that the *b*-axis is the most compressible axis in both phases, which is consistent with conclusions obtained from elastic constants calculations.

In order to describe the pressure (P) dependence of the volume (V) for the two phases we have used a third-order Birch−Murnaghan equation of state (EOS) [55]. The EOS determination was carried out using EosFit [56]. The obtained results for the zero-pressure volume ($V_0$), bulk modulus ($B_0$), and its first pressure derivative ($B_0'$) are given in Table 4. For the low-pressure phase we have made a fit for the quasi-hydrostatic regime (P < 10 GPa) and another including all the data of this phase up to the phase transition. The bulk-modulus of the two fits agree within error bars. The results of the fit using all data is shown in Fig. 4. The bulk modulus obtained from the DFT calculations using the EOS is comparable to the value obtained from the elastic constants, but 13% smaller than the one determined from experiments. Such differences are within the typical deviations between DFT calculations and experiments [57]. Similar differences have been found for $InNbO_4$ [15]. The bulk modulus we obtained when analyzing the quasi-hydrostatic regime, $B_0$ = 185(9) GPa, is similar with the same parameter determined for $InNbO_4$, $B_0$ = 179(2) GPa [15], and $BiNbO_4$, $B_0$ = 185(7) GPa [58]. However, $HoNbO_4$ is much less compressible than $GdNbO_4$, $B_0$ = 164(6) GPa [14], and $LaNbO_4$, $B_0$ = 111(3) GPa [10]. The differences observed between different lanthanide $RNbO_4$ niobates suggest that compressibility in these compounds should be dominated by the large $RO_8$ dodecahedra as observed in related compounds [59]. For the HP phase we found that the EOS parameters are very similar to those of the low-pressure phase. The values obtained from elastic constants are larger than those obtained from the other method because they were calculated at 21.9 GPa. A proper comparison should be done by obtaining the bulk modulus at the same pressure using the determined EOS (see table 4).

These values are shown in Table 4. Again DFT give a bulk modulus similar to elastic constants calculations. Both of them are 10% smaller than the value determined from experiments.

Before concluding the discussion of the influence of pressure in the crystal structure we would like to discuss in more detail the anisotropic compressibility of fergusonite $HoNbO_4$. Since the structure is monoclinic, the compressibility is described by a symmetric tensor with four elements different than zero [60]. The eigenvalues of the compressibility tensor will be the compressibility of the principal axes of compression (the eigenvectors) [60]. We have obtained them for $HoNbO_4$ considering only the data from the quasi-hydrostatic regime, using PASCAL [61]. Their values are given in Table 5. The major compression direction corresponds to the *b*-axis, as observed in Fig. 4 and deduced from elastic constants. The minimum compression direction is in the perpendicular plane making a 27º angle to the *a*-axis (from *a* to *c*). From the obtained eigenvalues the volume compressibility can be estimated as their sum, being $5.2(1) \times 10^{-3}$ $GPa^{-1}$, which corresponds to a bulk modulus of 192(9) GPa, which is in agreement with the result obtained from the EOS.

The infrared (IR) and Raman spectra of $HoNbO_4$ have been measured at ambient pressure. These spectra are shown in Fig. 5. According to group theory, the fergusonite structure has 36 vibrational modes [62]. Their representation at the zone centre of the Brillouin zone is: $\Gamma = 8A_g + 8A_u + 10B_g + 10B_u$. Among them, two $B_u$ and one $A_u$ modes are the acoustic modes. The remaining $A_u$ and $B_u$ modes (15 in total) are IR-active modes and the $A_g$ and $B_g$ modes (18 in total) are Raman-active modes. We have measured all these modes. They are indicated by red ticks in Fig. 5 and their frequencies are given in Tables 6 and 7. The present Raman spectrum fully agrees with the one reported by Siqueira *et al.* [63]. Results from our DFT calculations are reported in the same tables. There it can be seen that the agreement between calculations and experiments for Raman frequencies is quite good. In particular, in the last column of Table 6, we show the relative difference between frequencies ($R_\omega$) which is always smaller than 6%. For IR modes, with the exception of three low-frequency modes, calculations tend to underestimate the frequencies by less than 10% (see Table 7), which is typical for ternary oxides [64].

Calculations have helped to identify the symmetry of the different Raman and IR modes. This information is given in Tables 6 and 7. Calculations also contribute to the identification of the different vibrations associated to each mode. In the Raman spectrum there are four isolated modes at high frequency, they are related to internal stretching vibrations of the NbO$_4$ tetrahedron. Then, there are Raman modes in the 300–500 cm$^{-1}$ region which are mainly associated to bending vibrations on NbO$_4$. Finally, there are low-frequency modes, below 250 cm$^{-1}$, which basically involve movements of the niobate molecule as rigid units and of the Ho atoms. The IR modes can be also separated as internal vibrations of the NbO$_4$ tetrahedron at high frequencies, and external vibrations of it and vibrations of Ho at low frequencies. However, in the IR spectrum there is not a clear phonon gap as observed in the Raman spectrum.

From our simulations we have also obtained the pressure dependence of Raman and IR modes. The results are shown in Figs. 6 and 7. The pressure dependence of the modes is not linear. It can be described by a quadratic dependence $\omega(P) = \omega_0 + \alpha_1 P + \alpha_2 P^2$; which is valid for P < 20 GPa. The three parameters for each mode are given in Tables 6 and 7 to facilitate the comparison with future HP Raman and IR measurements. In the pressure range of stability of fergusonite-type HoNbO$_4$, all Raman modes harden under compression. The pressure coefficients are similar to those reported from HP Raman measurements on GdNbO$_4$ [14]. One of the most interesting features of the HP behaviour is the merging of some of the A$_g$ and B$_g$ modes. For instance, two of the high-frequency internal modes (See Fig. 6). Another fact to highlight is the existence of anti-crossing modes. In particular, the A$_g$ modes with frequencies close to 300 cm$^{-1}$ (represented in blue in the figure). In contrast with the Raman modes, we have observed that two IR modes soften under compression. These are the lowest frequency B$_u$ mode and one low-frequency A$_u$ mode (see Table 7 and Fig. 7). They are shown in green colour in the figure. The presence of soft modes is usually related with a weakening of the restoring force against the corresponding deformation associated to the phonon mode. This is connected with a collective instability that tends to make the crystal structure unstable [65]. This observation is consistent with the finding of a phase transition at 19 GPa. We have found that the B$_g$ soft mode becomes imaginary at 30 GPa. Therefore, the principal cause of the observed transition is the mechanical instability described before and not the phonon softening.

From our computer simulations we also obtain information on the phonons of the HP phase of HoNbO$_4$. According to group theory analysis, this structure has seventy-two vibrations being their representation: $\Gamma = 18A_g + 18A_u + 18B_g + 18B_u$. Among them one $A_u$ and two $B_u$ are the acoustic modes, $18A_g + 18B_g$ are the Raman-active modes, and $17A_u + 16B_u$ are the IR-active modes. The obtained frequencies and pressure dependences are represented in Fig. 8 and 9 and in Tables 8 and 9. As in the low-pressure phase, the pressure dependence of the Raman and IR modes of the HP phase follows a quadratic pressure dependence. The parameter describing these quadratic functions (valid for 20 GPa < P < 32 GPa) are listed in Tables 8 and 9. We found that most Raman modes harden under compression with only a few of them having negative pressure coefficients. Regarding the IR modes, there are five modes with negative pressure coefficients. We also observed several phonon crossings between $A_g$ and $B_g$ ($A_u$ and $B_u$) modes as can be seen in Figs. 8 and 9. Anti-crossing between modes of the same symmetry is also observed for both Raman and IR modes. A few examples are highlighted using blue colour in the figures. A last observation we have made is that in the HP phase we observed a reduction of the gap between internal stretching modes and the rest of the Raman-active modes. This is mainly caused by the increase of the number of modes. In particular, in the high-frequency region the HP phase has eight modes, while the low-pressure fergusonite phase has four modes. This change is a consequence of the increase of the coordination of Nb that happens at the phase transition as we discussed before.

**CONCLUDING REMARKS**

We have studied the HP behavior of HoNbO$_4$ by combining high-pressure XRD measurements and density-functional theory calculations. We have found evidence of the occurrence of a phase transition at 18.9(1.1) GPa. The crystal structure of high phase has been identified as monoclinic, belonging to space group P2$_1$/c. The phase transition takes place without detectable volume collapse and it is reversible. We have obtained information on the compressibility of two phases, being room-temperature equations of state determined. In particular, the bulk modulus of the fergusonite-type HoNbO$_4$ is 185(9) GPa being this compound one of the less compressible RNbO$_4$ orthoniobates. From calculations we also determined the elastic constants and moduli of the different phases of HoNbO$_4$. They also provide a description of the influence of pressure in Raman and IR phonons. The observed phase transition seems to be related to the violations of the Born stability criteria rather than the observed softening of one of IR mode. Therefore, mechanical instabilities are the fundamental mechanism behind the observed phase transition.

**Acknowledgements:** Dr. Alka B. Garg acknowledges Dr. Rakha Rao, Mr. Swayam Kesari, and Dr. Bobby Joseph for their help during ADXRD data collection at Express beamline of Elettra synchrotron source, Italy. The measurements were carried out during the beamtime allotted for proposal number 20165031. Dr. Garg Also acknowledges the Indian DST for financial support and Dr. Andrea Lausi for facilitating the beamtime. She also acknowledges Dr. Himal Bhatt for providing the ambient IR data. This work was supported by the Spanish Ministry of Science, Innovation and Universities under grants MAT2016-75586-C4-1/3-P and RED2018-102612-T (MALTA Consolider-Team network) and by Generalitat Valenciana under grant Prometeo/2018/123 (EFIMAT).

**FIGURES**

**Fig. 1**

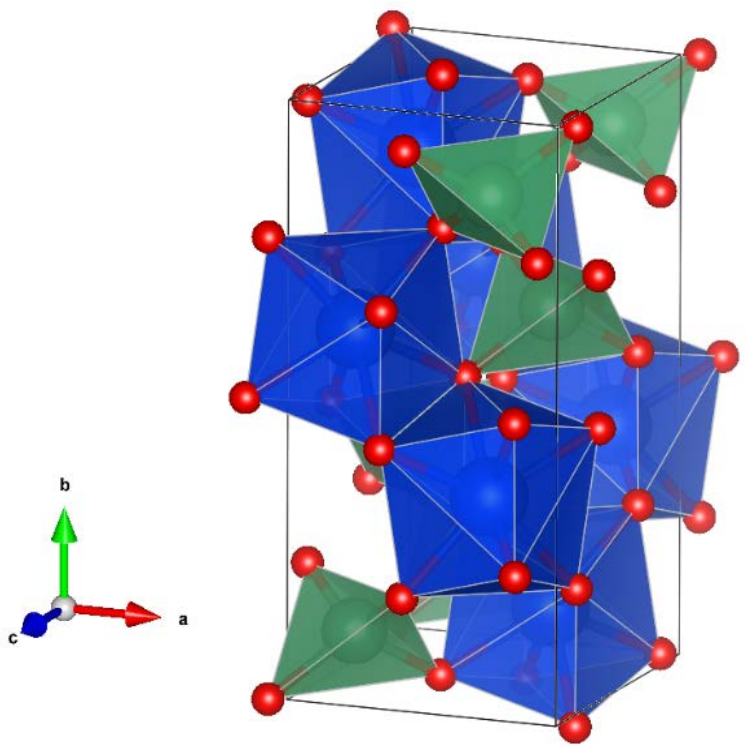

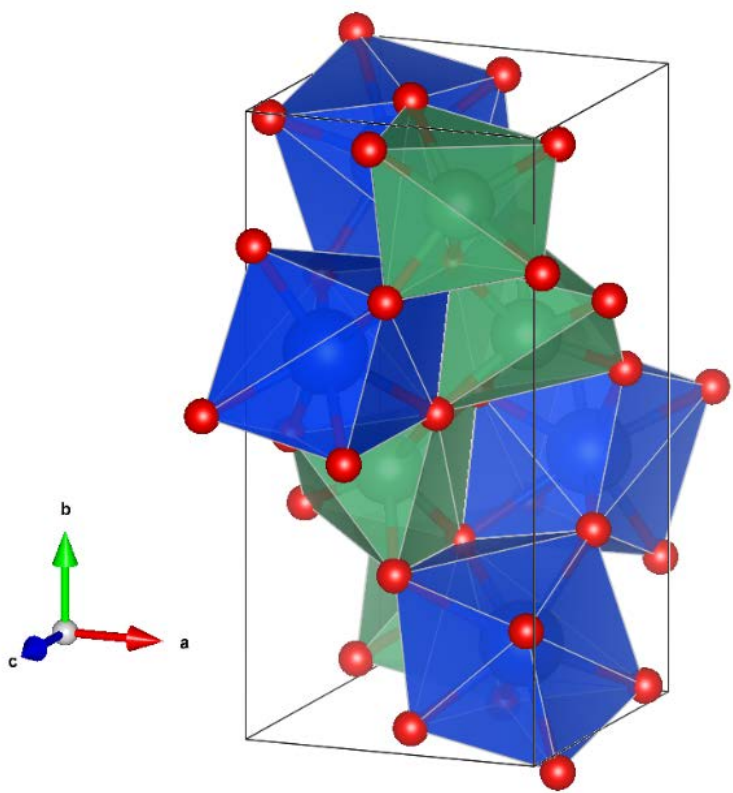

**Fig. 2**

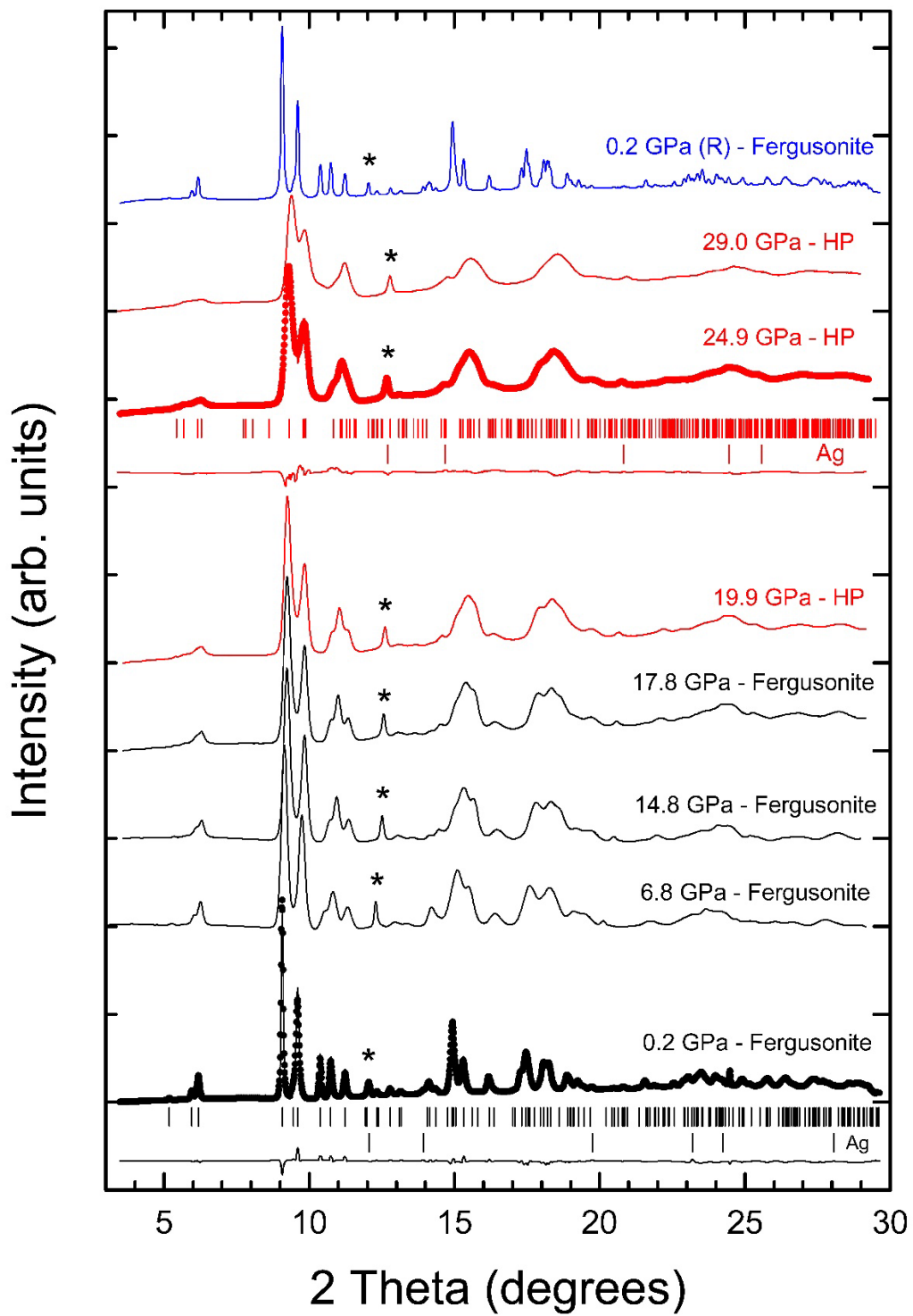

**Fig. 3**

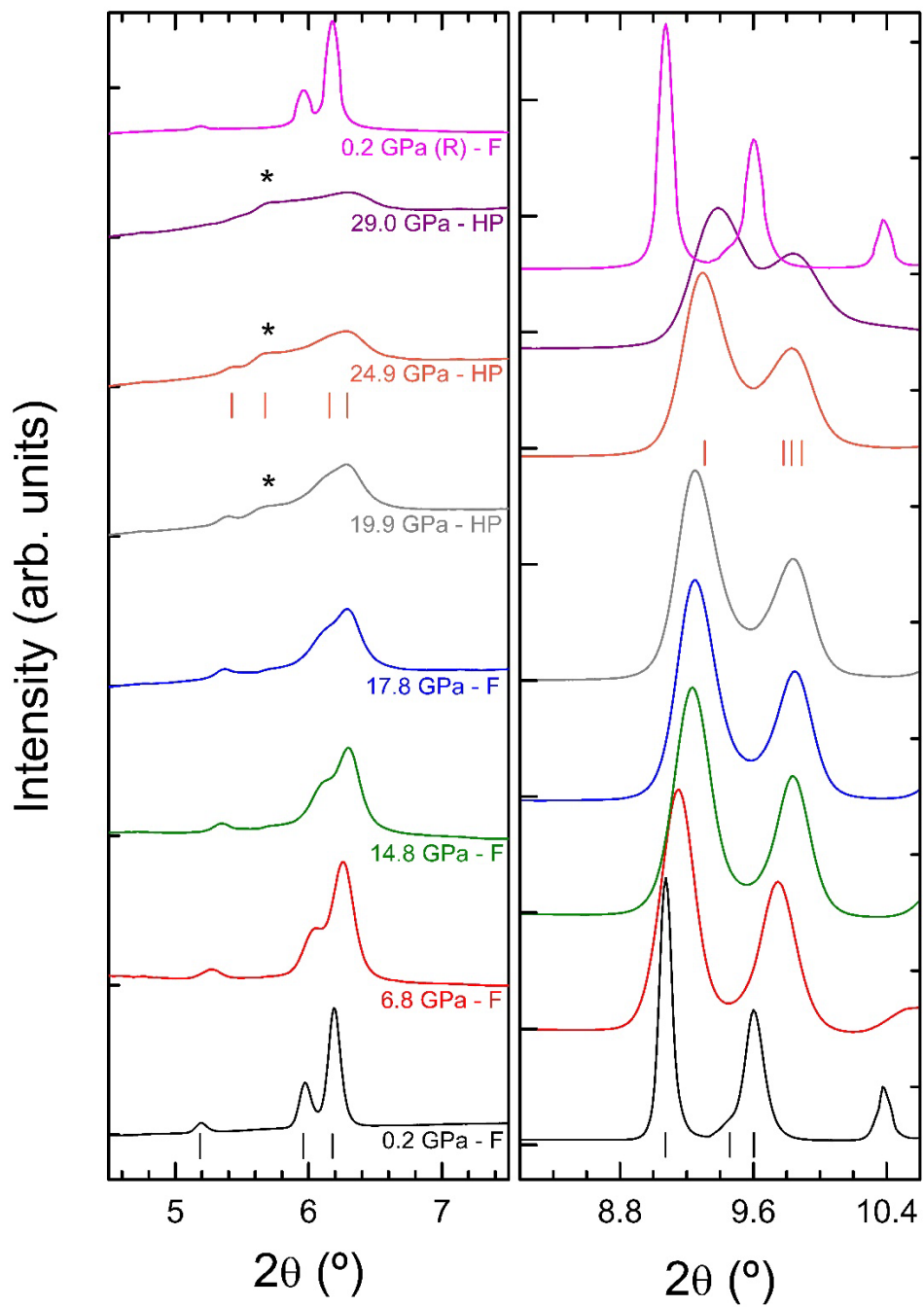

**Fig. 4**

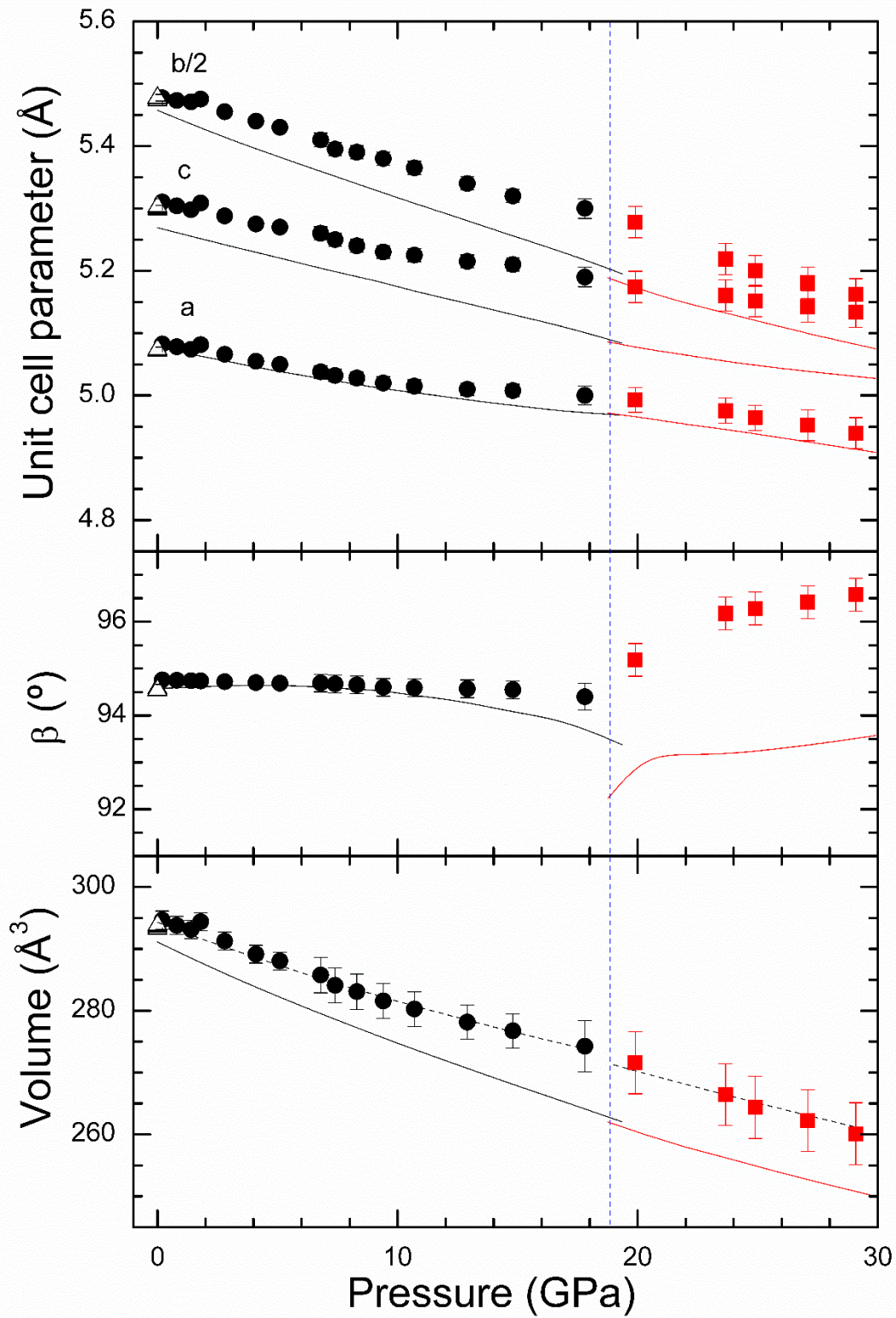

**Fig. 5**

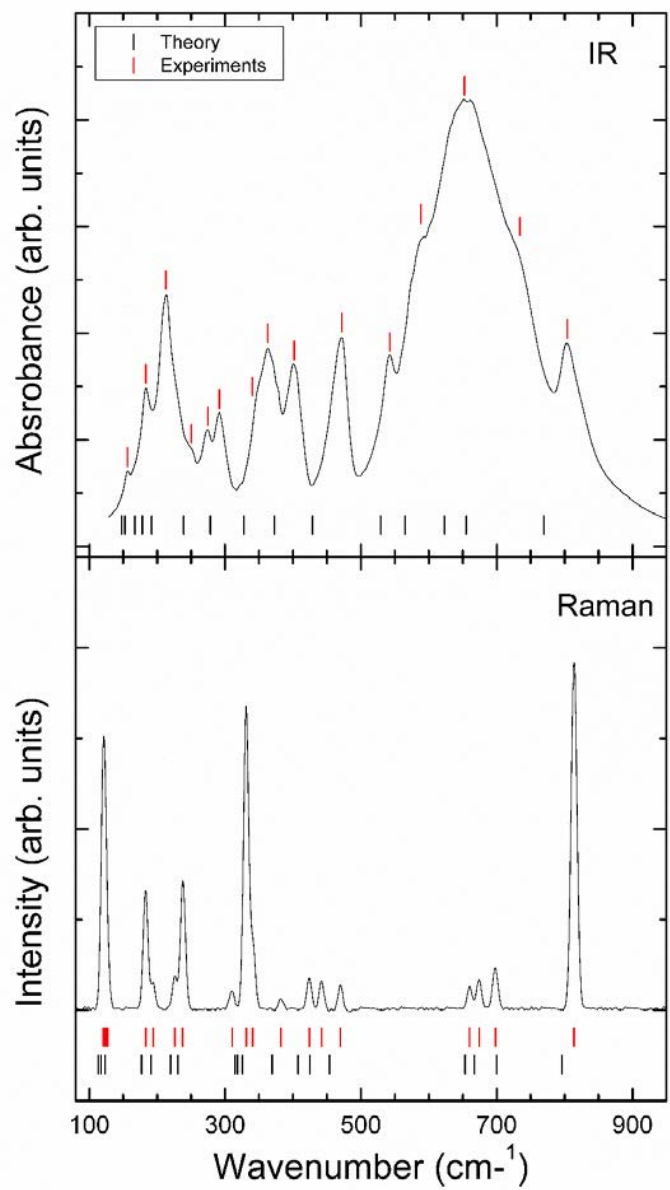

**Fig. 6**

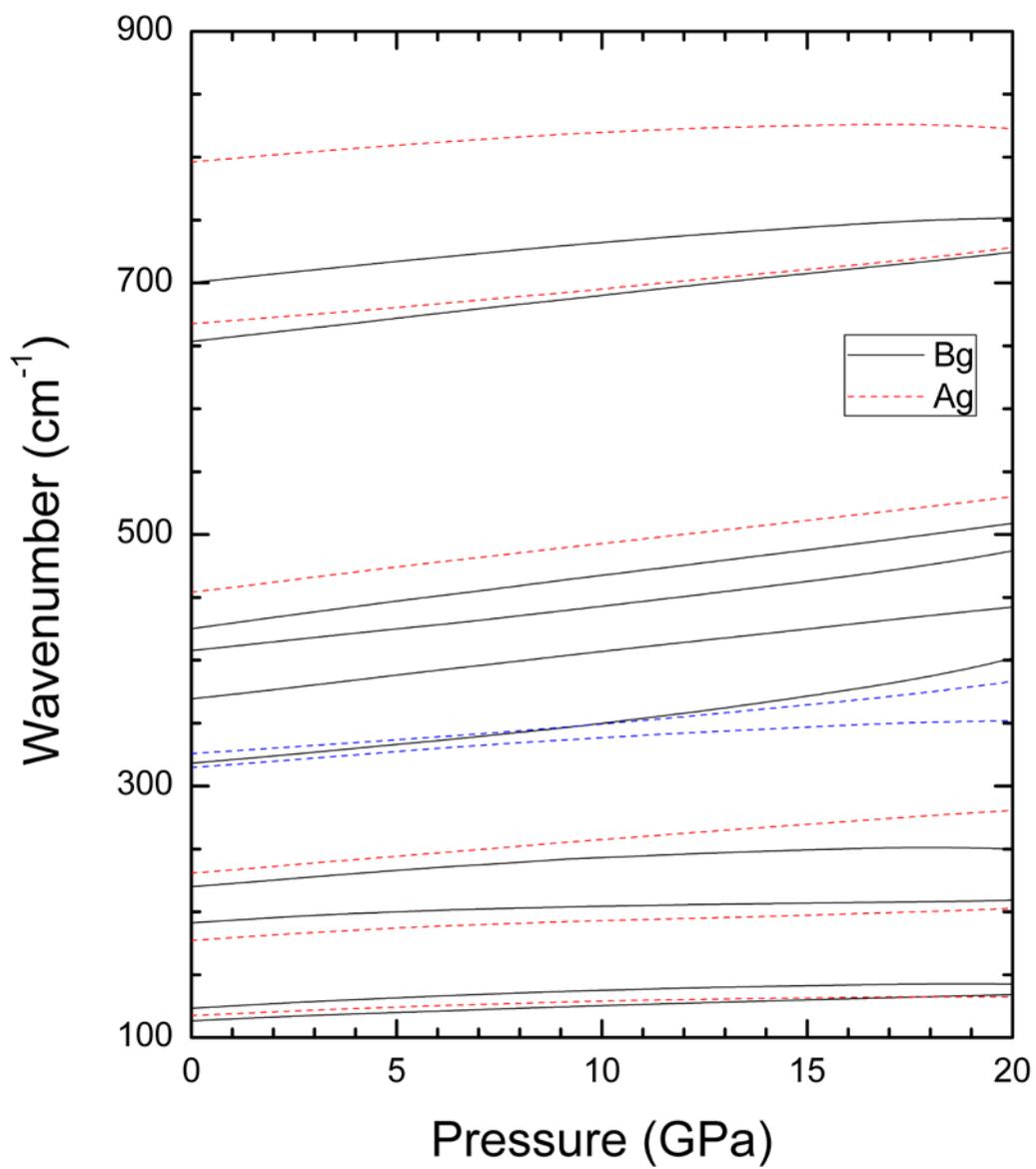

**Fig. 7**

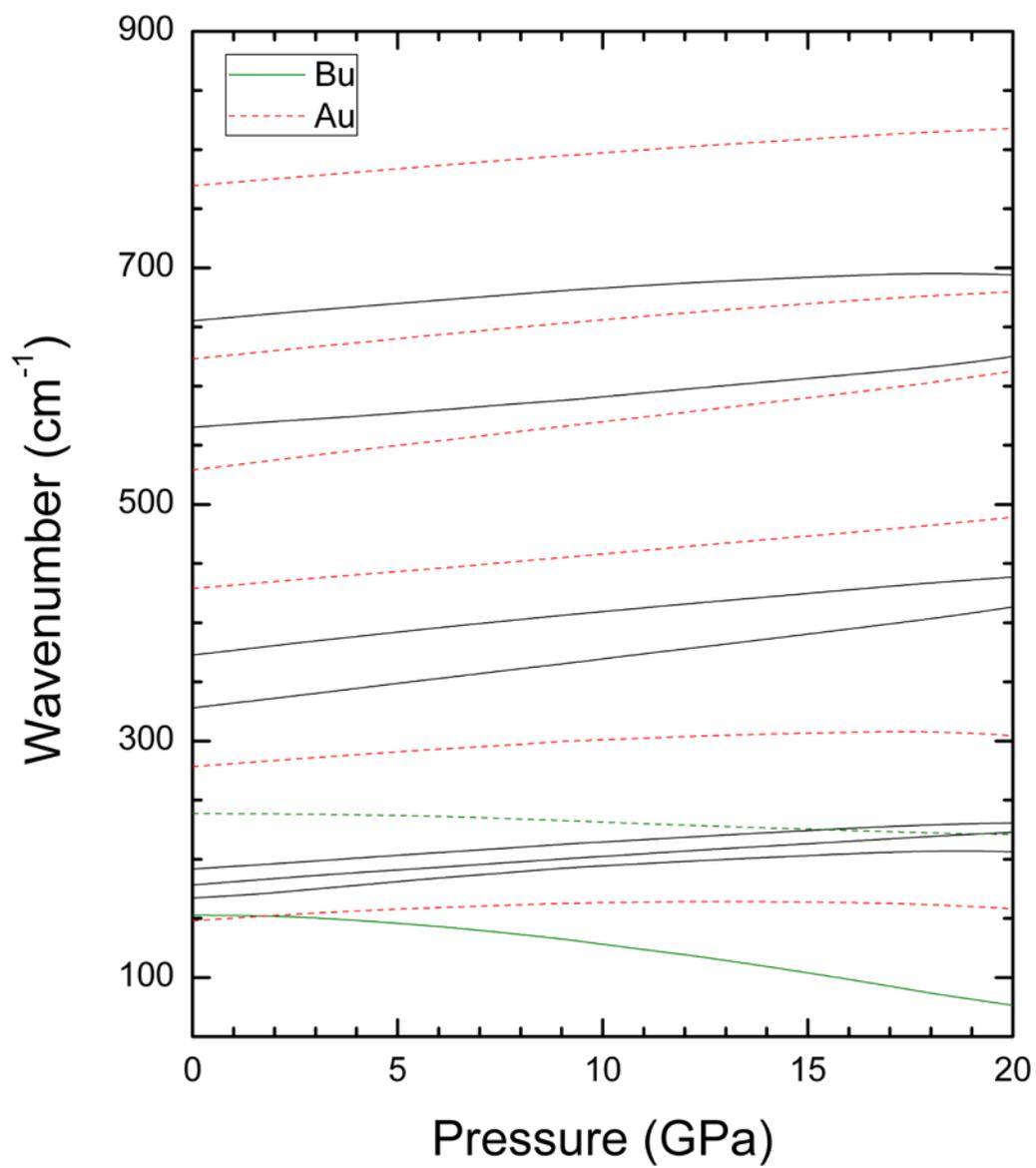

**Fig. 8**

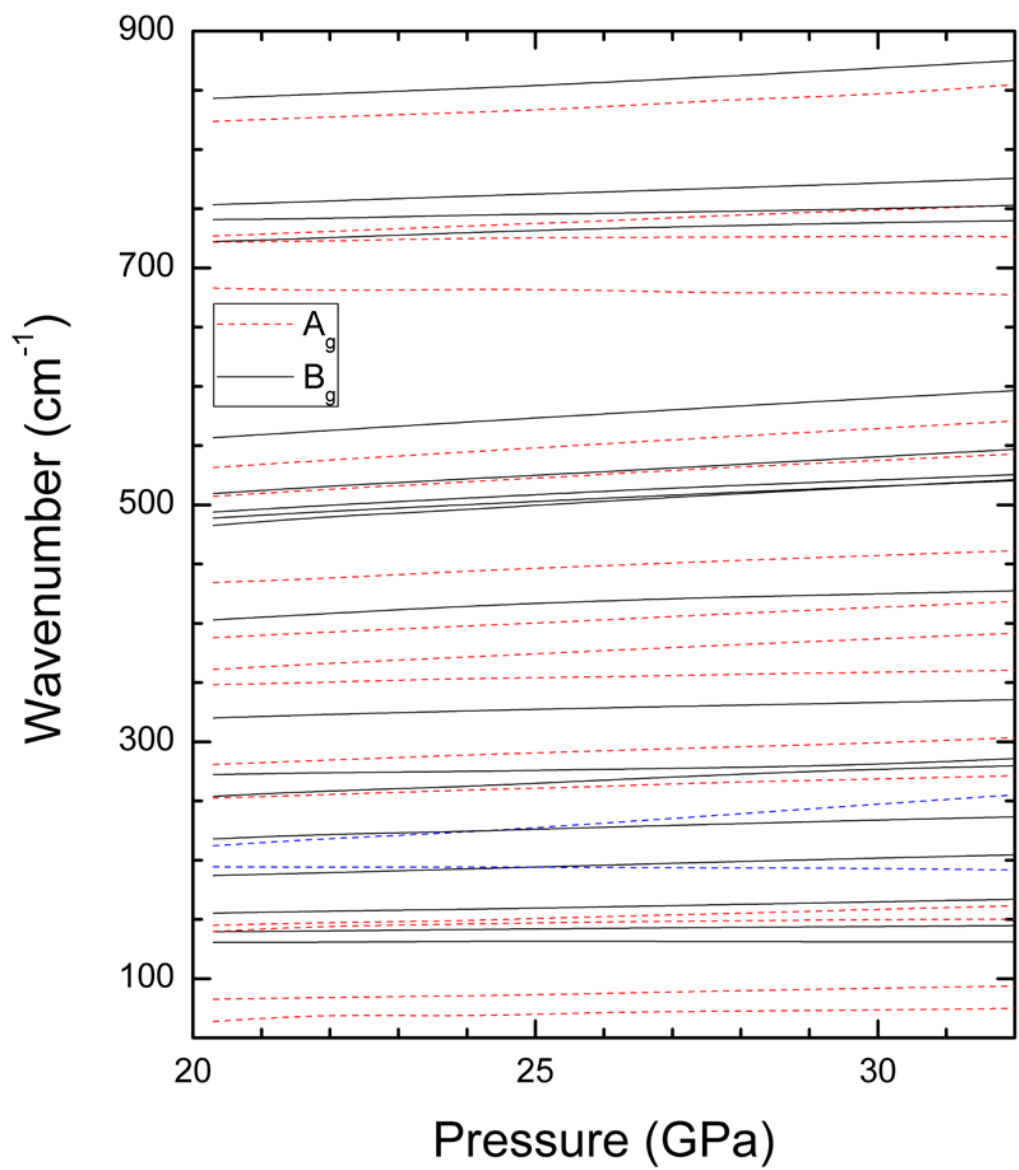

**Fig. 9**

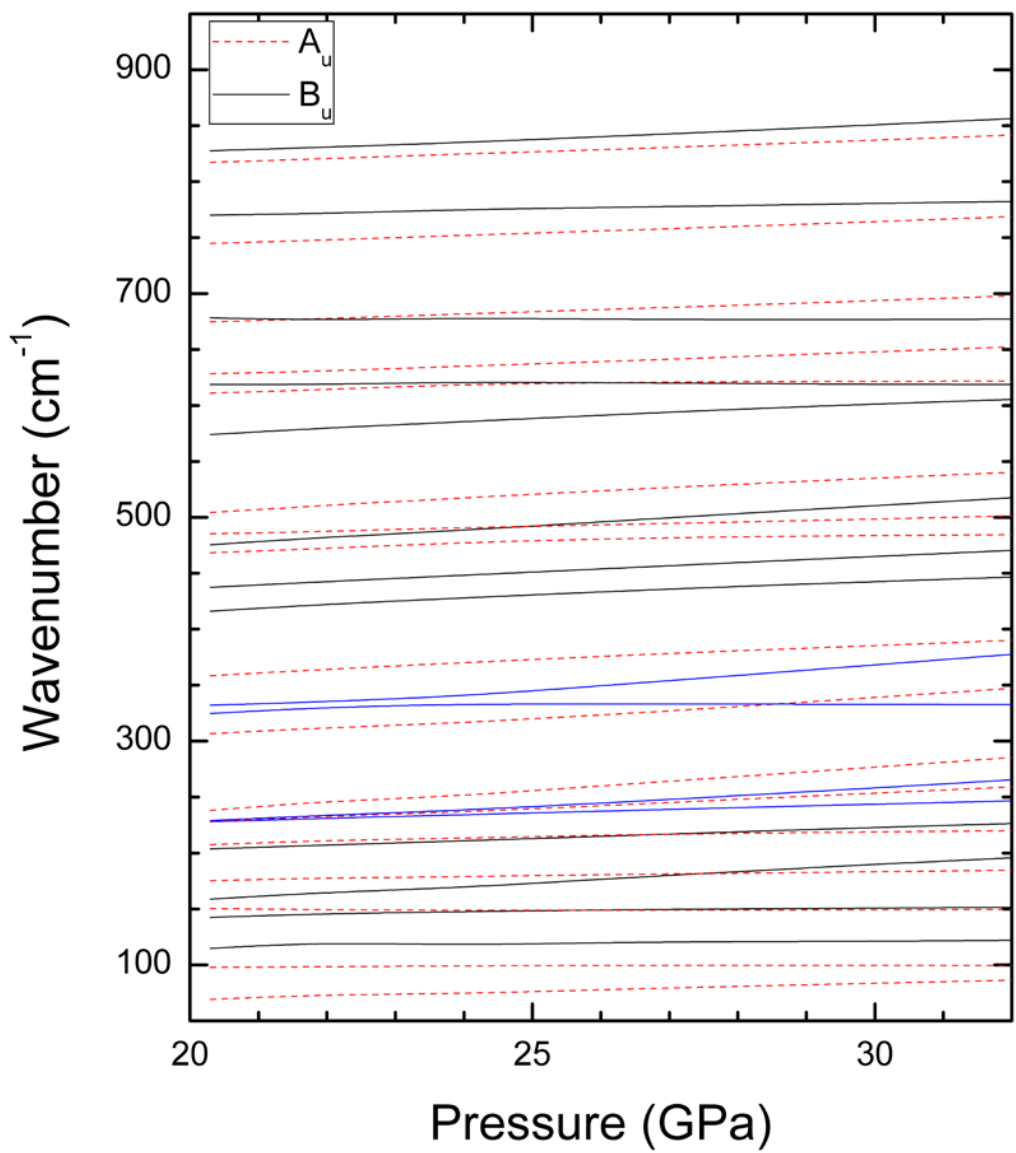

**FIGURE CAPTIONS**

**Fig. 1** Schematic representation of the crystal structure of fergusonite (top) and the HP phase (bottom). Ho and Nb coordination polyhedra are in blue and green color, respectively.

**Fig. 2** Selected powder XRD patterns from (a) 0.2 to 29.0 GPa. The identified phase is indicated. The Ag peak used for pressure determination is denoted by an asterisk. At 0.2 (24.9) GPa the experiments are show with symbols and the refinement for the fergusonite (HP) phase and residuals with lines. Ticks indicate the positions of calculated peaks. The XRD pattern measured upon pressure release (denoted by R) is shown to prove the reversibility of the phase transition.

**Fig.3** Zoom of two different angular regions two highlight changes observed in the XRD patterns. An asterisk is used to indicate the appearance of a peak at 19.9 GPa. In the two panels the same color corresponds to the same pressure which is indicated in the left panel. F indicates fergusonite, HP is used for the HP phase, and R indicates pressure release. Ticks indicate the position of Bragg peaks in the low- and high-pressure phases.

**Fig. 4.** Unit-cell parameters and volume vs. pressure. Black symbols are for the fergusonite phase. Red symbols for the HP phase. White triangles are ambient pressure results taken from the literature [3]. Solid lines are the results of DFT calculations. The black dashed lines are the equations of state described in the text. The vertical blue dashed line indicates the transition pressure.

**Fig. 5.** IR and Raman spectra of fergusonite-type $HoNbO_4$ at ambient conditions. Red ticks indicate the experimental frequencies and black ticks the theoretical frequencies.

**Fig. 6.** Calculated pressure dependence of Raman modes of fergusonite-type $HoNbO_4$. Dashed lines are used for $A_g$ modes and solid lines for $B_g$ modes. The blue color is used to facilitate the identification of anti-crossing modes.

**Fig. 7.** Calculated pressure dependence of IR modes of fergusonite-type $HoNbO_4$. Dashed lines are used for $A_u$ modes and solid lines for $B_u$ modes. The blue color is used to facilitate the identification of anti-crossing modes. Soft-modes are shown in green.

**Fig. 8.** Calculated pressure dependence of Raman modes of HP $HoNbO_4$. Dashed lines are used for $A_g$ modes and solid lines for $B_g$ modes. The blue color is used to facilitate the identification of anti-crossing modes.

**Fig. 9.** Calculated pressure dependence of IR modes of HP $HoNbO_4$. Dashed lines are used for $A_u$ modes and solid lines for $B_u$ modes. The blue color is used to facilitate the identification of anti-crossing modes.

# TABLES

## Table 1

| $a = 5.0777$ Å, $b = 10.9148$ Å, $c = 5.2692$ Å, and $\beta = 94.5733°$, space group I2/a | | | | |
|---|---|---|---|---|
| Atom | Site | x | y | z |
| Ho | 4e | 0.25 | 0.62887 | 0 |
| Nb | 4e | 0.25 | 0.10628 | 0 |
| $O_1$ | 8f | 0.24921 | 0.03117 | 0.03860 |
| $O_2$ | 8f | 0.16134 | 0.21179 | 0.49851 |
| $a = 4.9447$ Å, $b = 10.2608$ Å, $c = 5.0533$ Å, and $\beta = 93.1804°$, space group $P2_1/c$ | | | | |
| Atom | Site | x | y | z |
| Ho | 4e | 0.26760 | 0.87320 | 0.48160 |
| Nb | 4e | 0.26130 | 0.38290 | 0.50130 |
| $O_1$ | 4e | 0.03800 | 0.04170 | 0.25730 |
| $O_2$ | 4e | 0.53510 | 0.51030 | 0.73060 |
| $O_3$ | 4e | 0.41000 | 0.24850 | 0.27240 |
| $O_4$ | 4e | 0.07800 | 0.69880 | 0.60800 |

## Table 2

| Elastic constant | Fergusonite $HoNbO_4$ | HP Phase $HoNbO_4$ | Fergusonite $LaNbO_4$ | Fergusonite $YTaO_4$ |
|---|---|---|---|---|
| $C_{11}$ | 244.3 | 387.8 | 242 | 354.9 |
| $C_{12}$ | 83.1 | 147.8 | 83.8 | 112.3 |
| $C_{22}$ | 231.2 | 292.3 | 177 | 291.9 |
| $C_{13}$ | 151.9 | 194.9 | 185 | 146.1 |
| $C_{23}$ | 88.1 | 141.3 | 49.8 | 122.5 |
| $C_{33}$ | 272.1 | 383.4 | 212 | 343 |
| $C_{15}$ | 16.4 | -35.3 | 4.1 | 1.0 |
| $C_{25}$ | -4.75 | -2.7 | -16.4 | –20.3 |
| $C_{35}$ | -21.4 | 15.2 | 29.3 | –57.5 |
| $C_{55}$ | 89 | 30.8 | 11.8 | 82.8 |
| $C_{44}$ | 55.7 | 44.6 | 43.9 | 88.1 |
| $C_{46}$ | -3.6 | 7.0 | -8.9 | –1.6 |
| $C_{66}$ | 60.4 | 51.9 | 53.6 | 73.5 |

**Table 3**

| Phase | Approximation | $B_0$ | E | G | $\nu$ | $B_0/G$ | $A_U$ | $H_V$ |
|---|---|---|---|---|---|---|---|---|
| Fergusonite | Voigt | 154.9 | 180.9 | 69.3 | 0.305 | 2.23 | 0.49 | 6.94 |
|  | Reuss | 150.7 | 166.8 | 63.4 | 0.315 | 2.38 |  |  |
|  | Hill | 152.8 | 173.9 | 66.3 | 0.310 | 2.30 |  |  |
| HP phase | Voigt | 225.7 | 175.6 | 64.2 | 0.370 | 3.51 | 1.99 | 3.23 |
|  | Reuss | 216.4 | 128.7 | 45.9 | 0.401 | 4.71 |  |  |
|  | Hill | 221.1 | 152.4 | 55.0 | 0.385 | 4.02 |  |  |

**Table 4**

| Phase | Conditions | $V_0$ (Å$^3$) | $B_0$ (GPa) | $B'_0$ |
|---|---|---|---|---|
| Fergusonite | Theory (EOS) | 290 | 161 | 2.8 |
|  | Quasi-hydrostatic | 294.3(9) | 185(5) | 4.2(8) |
|  | P ≤ 17.8 GPa | 294.3(9) | 195(9) | 6.5(9) |
|  | Theory (Elastic constants) |  | 150.7 – 154.9 |  |
|  | Compressibility Tensor |  | 191 |  |
| HP phase | Theory (EOS) | 288 | 169 | 3.2 |
|  |  |  | 235* |  |
|  | 19.9 GPa < 29.1 GPa | 296(2) | 187(9) | 3.5(9) |
|  |  |  | 259(15)* |  |
|  | Theory (Elastic constants) |  | 216.4 – 225.7 |  |

**Table 5**

| | |
|---|---|
| $\lambda_1 = 2.2(1)\ 10^{-3}$ GPa$^{-1}$ | $e_{v1} = (0, 1, 0)$ |
| $\lambda_2 = 1.7(1)\ 10^{-3}$ GPa$^{-1}$ | $e_{v2} = (-0.4402, 0, 0.8979)$ |
| $\lambda_3 = 1.3(1)\ 10^{-3}$ GPa$^{-1}$ | $e_{v3} = (-0.8914, 0, -0.4533)$ |

**Table 6**

| Mode | Experiment | Theory | | | $R_\omega$ |
|---|---|---|---|---|---|
| | $\omega_0$ (cm$^{-1}$) | $\omega_0$ (cm$^{-1}$) | $\alpha_1$ (cm$^{-1}$/GPa) | $\alpha_2$ (cm$^{-1}$/GPa$^2$) | |
| B$_g$ | 121 | 113.3 | 1.32 | -0.015 | 0.0636 |
| A$_g$ | 124 | 117.4 | 1.55 | -0.040 | 0.0532 |
| B$_g$ | 127 | 123.3 | 1.93 | -0.048 | 0.0291 |
| A$_g$ | 183 | 177.2 | 1.69 | -0.024 | 0.0317 |
| B$_g$ | 194 | 191.2 | 1.58 | -0.037 | 0.0144 |
| B$_g$ | 226 | 220.0 | 3.27 | -0.086 | 0.0265 |
| A$_g$ | 238 | 230.6 | 2.90 | -0.019 | 0.0311 |
| A$_g$ | 311 | 314.6 | 2.99 | -0.054 | -0.0116 |
| B$_g$ | 331 | 318.2 | 1.84 | 0.111 | 0.0387 |
| A$_g$ | 341 | 325.7 | 1.73 | 0.055 | 0.0449 |
| B$_g$ | 382 | 369.2 | 3.92 | -0.013 | 0.0335 |
| B$_g$ | 424 | 407.5 | 2.89 | 0.051 | 0.0389 |
| B$_g$ | 442 | 425.0 | 4.20 | -0.002 | 0.0385 |
| A$_g$ | 470 | 454.0 | 3.90 | -0.006 | 0.0340 |
| B$_g$ | 660 | 653.3 | 3.78 | -0.011 | 0.0101 |
| A$_g$ | 674 | 667.4 | 2.46 | 0.028 | 0.0098 |
| B$_g$ | 699 | 700.2 | 3.96 | -0.067 | -0.0017 |
| A$_g$ | 814 | 796.1 | 3.63 | -0.111 | 0.0220 |

**Table 7**

| Mode | Experiment | Theory | | | $R_\omega$ |
|---|---|---|---|---|---|
| | $\omega_0$ (cm$^{-1}$) | $\omega_0$ (cm$^{-1}$) | $\alpha_1$ (cm$^{-1}$/GPa) | $\alpha_2$ (cm$^{-1}$/GPa$^2$) | |
| A$_u$ | 156 | 148.0 | 2.64 | -0.105 | 0.0513 |
| B$_u$ | 183 | 152.6 | -1.70 | -0.108 | 0.1661 |
| B$_u$ | 213 | 167.3 | 2.60 | -0.018 | 0.2145 |
| B$_u$ | 250 | 178.2 | 2.77 | -0.038 | 0.2872 |
| B$_u$ | 275 | 191.8 | 3.79 | -0.086 | 0.3025 |
| A$_u$ | 292 | 238.7 | -0.74 | -0.011 | 0.1825 |
| A$_u$ | 340 | 278.4 | 3.50 | -0.105 | 0.1812 |
| B$_u$ | 363 | 327.9 | 3.93 | 0.016 | 0.0967 |
| B$_u$ | 402 | 372.7 | 4.06 | -0.038 | 0.0729 |
| A$_u$ | 472 | 428.9 | 4.06 | -0.038 | 0.0913 |
| A$_u$ | 542 | 529.1 | 3.83 | 0.015 | 0.0238 |
| B$_u$ | 588 | 565.2 | 2.04 | 0.046 | 0.0388 |
| A$_u$ | 653 | 623.1 | 3.88 | -0.050 | 0.0458 |
| B$_u$ | 734 | 655.3 | 3.84 | -0.091 | 0.1072 |
| A$_u$ | 804 | 769.4 | 3.22 | -0.038 | 0.0430 |

**Table 8**

| Mode | ω (cm$^{-1}$) | ω$_0$ (cm$^{-1}$) | α$_1$ (cm$^{-1}$/GPa) | α$_2$ (cm$^{-1}$/GPa$^2$) |
|---|---|---|---|---|
| A$_g$ | 63.7 | 19.8 | 3.13 | -0.044 |
| A$_g$ | 82.6 | 32.1 | 3.30 | -0.044 |
| B$_g$ | 130.7 | 92.9 | 1.60 | 0.015 |
| B$_g$ | 139.6 | 122.0 | 1.12 | -0.018 |
| A$_g$ | 139.8 | 122.4 | 1.09 | -0.012 |
| A$_g$ | 145.0 | 133.7 | 0.02 | 0.026 |
| B$_g$ | 155.2 | 142.1 | 0.44 | 0.011 |
| B$_g$ | 186.9 | 149.9 | 2.01 | -0.010 |
| A$_g$ | 194.6 | 181.1 | 1.19 | -0.027 |
| A$_g$ | 212.1 | 175.5 | 0.64 | 0.058 |
| B$_g$ | 217.9 | 172.1 | 2.73 | -0.022 |
| A$_g$ | 252.4 | 199.2 | 3.25 | -0.031 |
| B$_g$ | 254.0 | 184.4 | 4.17 | -0.037 |
| B$_g$ | 272.2 | 297.2 | -2.63 | 0.070 |
| A$_g$ | 280.7 | 232.5 | 2.70 | -0.015 |
| B$_g$ | 320.0 | 275.1 | 2.81 | -0.029 |
| A$_g$ | 348.2 | 309.3 | 2.48 | -0.028 |
| A$_g$ | 361.1 | 288.5 | 4.21 | -0.031 |
| A$_g$ | 387.8 | 330.1 | 2.99 | -0.007 |
| B$_g$ | 402.8 | 281.5 | 8.45 | -0.122 |
| B$_g$ | 434.5 | 362.1 | 4.32 | -0.038 |
| A$_g$ | 482.8 | 391.5 | 5.31 | -0.039 |
| B$_g$ | 488.9 | 409.2 | 4.73 | -0.039 |
| B$_g$ | 493.9 | 400.1 | 5.86 | -0.061 |
| A$_g$ | 507.2 | 421.9 | 4.95 | -0.036 |
| B$_g$ | 509.4 | 438.2 | 3.76 | -0.011 |
| A$_g$ | 531.5 | 447.8 | 4.61 | -0.024 |
| B$_g$ | 556.6 | 474.1 | 4.49 | -0.021 |
| A$_g$ | 683.0 | 660.0 | 0.44 | -0.016 |
| A$_g$ | 721.8 | 680.0 | 3.08 | -0.051 |
| B$_g$ | 722.2 | 646.2 | 5.14 | -0.069 |
| A$_g$ | 727.1 | 673.8 | 2.86 | -0.012 |
| B$_g$ | 740.9 | 730.2 | 0.21 | 0.015 |
| B$_g$ | 753.4 | 715.0 | 1.90 | 0.000 |
| A$_g$ | 823.7 | 823.0 | -1.57 | 0.080 |
| B$_g$ | 843.2 | 827.5 | -0.45 | 0.061 |

**Table 9**

| Mode | ω (cm$^{-1}$) | ω$_0$ (cm$^{-1}$) | α$_1$ (cm$^{-1}$/GPa) | α$_2$ (cm$^{-1}$/GPa$^2$) |
|---|---|---|---|---|
| A$_u$ | 69.1 | 39.8 | 1.51 | -0.001 |
| A$_u$ | 97.8 | 79.6 | 1.37 | -0.023 |
| B$_u$ | 114.7 | 83.4 | 2.33 | -0.035 |
| B$_u$ | 142.5 | 88.0 | 3.97 | -0.062 |
| A$_u$ | 150.4 | 170.5 | -1.61 | 0.030 |
| B$_u$ | 158.7 | 99.8 | 2.74 | 0.008 |
| A$_u$ | 175.2 | 145.9 | 1.89 | -0.022 |
| B$_u$ | 203.7 | 159.5 | 2.33 | -0.008 |
| A$_u$ | 207.3 | 145.7 | 4.33 | -0.063 |
| B$_u$ | 228.0 | 191.2 | 1.96 | -0.007 |
| A$_u$ | 228.1 | 192.7 | 1.19 | 0.027 |
| B$_u$ | 229.0 | 207.3 | -0.21 | 0.063 |
| A$_u$ | 238.1 | 184.0 | 1.87 | 0.040 |
| A$_u$ | 306.6 | 295.2 | -1.25 | 0.090 |
| B$_u$ | 324.8 | 233.2 | 7.13 | -0.126 |
| B$_u$ | 332.1 | 340.9 | -3.30 | 0.139 |
| A$_u$ | 358.5 | 271.3 | 5.33 | -0.050 |
| B$_u$ | 416.1 | 320.4 | 6.09 | -0.067 |
| B$_u$ | 437.4 | 371.8 | 3.49 | -0.013 |
| A$_u$ | 468.4 | 357.0 | 8.08 | -0.128 |
| B$_u$ | 475.6 | 406.0 | 3.36 | 0.004 |
| A$_u$ | 485.2 | 454.9 | 1.57 | -0.004 |
| A$_u$ | 504.4 | 407.2 | 5.89 | -0.054 |
| B$_u$ | 574.0 | 481.7 | 5.75 | -0.059 |
| A$_u$ | 610.9 | 513.1 | 7.31 | -0.123 |
| B$_u$ | 618.7 | 587.3 | 2.53 | -0.048 |
| A$_u$ | 628.3 | 599.6 | 0.97 | 0.021 |
| A$_u$ | 674.7 | 641.5 | 1.41 | 0.011 |
| B$_u$ | 678.2 | 687.1 | -0.72 | 0.013 |
| A$_u$ | 744.8 | 713.0 | 1.25 | 0.015 |
| B$_u$ | 770.0 | 731.4 | 2.41 | -0.025 |
| A$_u$ | 817.2 | 784.3 | 1.32 | 0.014 |
| B$_u$ | 827.5 | 805.1 | 0.22 | 0.043 |

**TABLE CAPTIONS**

**Table 1**. DFT calculated crystal structure of fergusonite-type $HoNbO_4$ at ambient pressure and the monoclinic HP phase at 23.7 GPa.

**Table 2.** Elastic constants (GPa) of fergusonite and monoclinic HP $HoNbO_4$ calculated by first principles. The results for fergusonite corresponds to ambient pressure and the results for the HP phase for 21.9 GPa. The results for fergusonite $LaNbO_4$ [41] and $YTaO_4$ [42] are included for comparison.

**Table 3.** Calculated bulk modulus ($B_0$), shear modulus (G), and Young modulus (E) given in GPa for the two phase of $HoNbO_4$. For fergusonite the results are at ambient pressure and for the monoclinic HP phase at 21.9 GPa. The Poisson's ratio ($\nu$), K/G ratio, and universal anisotropy index ($A_U$), which are dimensionless, are also reported. The last column shows the Vickers Hardness (in GPa).

**Table 4.** Equation of state parameters and bulk modulus determined from different methods. For the HP phase the bulk modulus determined from the elastic constants has been calculated at 21.9 GPa. For comparison the value of the bulk modulus at 21.9 GPa, obtained by using the determined experimental and theoretical EOS, is also given and denoted by *.

**Table 5.** Eigenvalues, $\lambda_i$, and eigenvectors, $e_{vi}$, of the isothermal compressibility tensor of fergusonite-type $HoNbO_4$.

**Table 6.** Experimental and calculated Raman frequencies ($\omega_0$) of fergusonite-type $HoNbO_4$ at ambient pressure including mode assignment. The relative difference between measured and calculated frequencies is also given ($R_\omega$). The linear ($\alpha_1$) and quadratic ($\alpha_2$) coefficients of the calculated pressure dependence are also reported.

**Table 7.** Ambient-pressure experimental and calculated IR frequencies ($\omega_0$) of fergusonite-type $HoNbO_4$ including mode assignment. The relative difference between measured and calculated frequencies is also given ($R_\omega$). The linear ($\alpha_1$) and quadratic ($\alpha_2$) coefficients of the calculated pressure dependence are also reported.

**Table 8.** Calculated Raman frequencies ($\omega$) of HP $HoNbO_4$ at 20.3 GPa. The independent ($\omega_0$), linear ($\alpha_1$), and quadratic ($\alpha_2$) coefficients of the calculated pressure dependence are also reported.

**Table 9.** Calculated Raman frequencies ($\omega$) of HP $HoNbO_4$ at 20.3 GPa. The independent ($\omega_0$), linear ($\alpha_1$), and quadratic ($\alpha_2$) coefficients of the calculated pressure dependence are also reported.